\documentclass{IEEEtran}
\usepackage{float}
\usepackage{soul}
\usepackage{mathtools}
\usepackage{epsfig,wrapfig}
\usepackage{epstopdf}
\usepackage{fancyhdr}
\usepackage{psfrag}
\usepackage{lipsum}
\usepackage{xcolor}
\usepackage{multirow}
\usepackage{booktabs}
\usepackage{rotfloat}
\usepackage{cite}
\usepackage{amsmath}
\usepackage{cite}
\usepackage{graphicx}
\usepackage{esint}
\usepackage{tablefootnote}
\usepackage{hhline}
\usepackage{nicematrix}
\usepackage[normalem]{ulem}

\newcommand\blueout{\bgroup\markoverwith
{\textcolor{blue}{\rule[.5ex]{2pt}{0.8pt}}}\ULon}
\usepackage{efbox,graphicx}
\efboxsetup{linecolor=blue,linewidth=1.5pt}

\newcommand{\E}{\mathbf{E}}
\renewcommand{\H}{\mathbf{H}}
\newcommand{\D}{\mathbf{D}}
\newcommand{\B}{\mathbf{B}}
\renewcommand{\S}{\mathbf{S}}
\newcommand{\g}{\mathbf{g}}
\newcommand{\mat}[1]{\overline{\overline{#1}}}
\renewcommand{\r}{\mathrm{r}}
\renewcommand{\t}{\mathrm{t}}
\renewcommand{\d}{\mathrm{d}}

\begin{document}

\title{Wave-Medium Interactions in Dynamic \\ Matter and Modulation Systems}
\author{Zhiyu Li, Xikui Ma, Zo{\'e}-Lise Deck-L{\'e}ger, Amir Bahrami, and Christophe Caloz
\thanks{Zhiyu Li and Xikui Ma are with Department of Electrical Engineering, Xi’an Jiaotong University, Xi’an, 710049, China (e-mail: lizhiyu@stu.xjtu.edu.cn). }
\thanks{Zo{\'e}-Lise Deck-L{\'e}ger is with Department of Electrical Engineering, Polytechnique Montr{\'e}al, Montr{\'e}al, H3T 1J4, Canada.}
\thanks{Amir Bahrami and Christophe Caloz are with Department of Electrical Engineering, KU Leuven, Leuven, 3001, Belgium.}}

\IEEEtitleabstractindextext{%
\begin{abstract}
Space-time modulation systems have recently garnered significant attention due to their resemblance to moving-matter systems, unique properties and promising applications. Unlike conventional moving-matter systems, modulation systems do not involve any net motion of matter, and are therefore easier to implement and capable to attain relativistic and superluminal velocities. However, the fundamental wave-medium interaction aspects in such media, such as scattering and energy-momentum relations, have been essentially unexplored to date. In this paper, we fill this gap, considering three dynamic systems: moving-matter blocs, moving-perturbation interfaces and moving-perturbation truncated periodic structures, and provide corresponding general formulations along with comparisons. Our investigation reveals significant roles played by the ``pushing'' and ``pulling'' effects of the moving interface onto the wave in such systems. Moreover, it describes different energy and momentum transfers between moving media and homogenized moving-perturbation structures that result from conventional and reverse Fresnel-Fizeau drag effects.
\end{abstract}

}
\maketitle
\IEEEdisplaynontitleabstractindextext
\IEEEpeerreviewmaketitle


\section{Introduction}\label{sec:introduction}

Dynamic systems encompass not only \emph{moving-matter systems}~\cite{Tai_1964_study,Ostrovskii_1975_some,Leonhardt_1999_optics,Leonhardt_2000_relativistic,Grzegorczyk_2006_electrodynamics,Deck_2021_electromagnetic}, which involve moving atoms and molecules, but also \emph{moving-perturbation (or modulation) systems}~\cite{Lurie_2007_mathematical,Biancalana_2007_dynamics,Caloz_2019_spacetime1,Caloz_2019_spacetime2,Yin_2022_floquet,Caloz_GSTEMs}, whose atoms and molecules oscillate without any net motion about their bound position in the medium, and where the moving entities are perturbation interfaces between different modulated media. This modulation may manifest as single step or gradient interfaces~\cite{Morgenthaler_1958_velocity,Felsen_1970_wave,Hadad_2020_soft,Rizza_2022_SPM,Liberal_2023_quantum,Bahrami_2023_electrodynamics}, or periodic interfaces~\cite{Morgenthaler_1958_velocity,Cassedy_1963_dispersion1,Reed_2003_color,Deck_2019_uniform,Sharabi_2021_diordered}. The spatial and temporal variation of electromagnetic parameters (e.g., refractive index) within these dynamic systems breaks time-reversal symmetry~\cite{Caloz_2018_nonreciprocity} and modifies the conservation laws of momentum and energy~\cite{Noether_1918_invariante,Caloz_2019_spacetime2,Francesco_2022_energy,Ortega_2023_tutorial}, leading to the emergence of new physics and applications. Among these are time refocusing~\cite{Bacot_2016_timereversal,Deck_2018_wave}, frequency transitions~\cite{Tsai_1967_wave,Taravati_2022_microwave}, magnetless nonreciprocity~\cite{Yu_2009_opticalisolation,Correas_2016_NonreciGraphene,Taravati_2022_low}, parametric amplification~\cite{Tien_1958_parametric,Galiffi_2019_broadband,Galiffi_2022_archimedes}, space-time homogenization~\cite{Pacheco_2020_effective,Huidobro_2021_homogenization,Serra_2023_homogenization,Prudencio_2023_replicating,Engheta_2023_4D}, inverse prism and temporal rainbow scattering~\cite{Akbarzadeh_2018_inverse,Stefanini_2023_temporal}, temporal double-slit diffraction~\cite{Tirole_2023_double}, linear time-invariant limits breaking~\cite{Shlivinski_2018_BF,Hayran_2023_templimitations} and extreme energy accumulation~\cite{Mirmoosa_2019_timereactive}.

Understanding wave interactions in such systems, particularly the associated energy-momentum relations, is vital for both academic research and practical applications. While studies have explored specific cases, such as treating the first medium as a vacuum in moving-matter systems~\cite{Costen_1965_three,Daly_1967_energy}, solutions for moving-perturbation systems remain elusive. A comprehensive, general solution to these canonical problems is essential for advancing our understanding and unlocking the full potential of dynamic systems across diverse fields of science and engineering.

We present here a thorough study on energy-momentum relations in dynamic matter and modulation systems. First, we demonstrate and compare the structure of different systems (Sec.~\ref{sec:types}). Then, we generalize the surface power transfer and force density formulas initially derived for moving-matter systems to encompass all dynamic systems (Sec.~\ref{sec:energy}). Next, we provide analytical solution for harmonic incidence (Sec.~\ref{sec:harmonic}). Afterward, we perform an in-depth energy-momentum analysis, shedding light on key phenomena such as the ``pushing'' and ``pulling'' effects of the moving interface onto the wave, and the Fresnel-Fizeau drag of the media; we also validate our theory by full-wave Finite-Difference Time-Domain (FDTD) simulations (Sec.~\ref{sec:analysis}). Finally, we close the paper with a conclusion (Sec.~\ref{sec:conclusion}).

\section{Types of Dynamic Systems}\label{sec:types}
\begin{figure}[ht!]
    \centering
    \includegraphics[width=8.6cm]{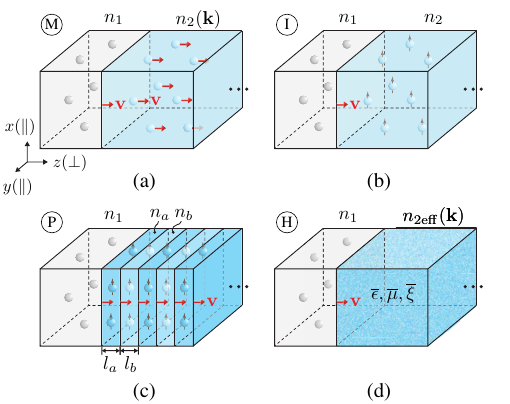}
    \caption{\label{fig:illustration} Dynamic systems considered in the paper. (a)~Moving-matter bloc (M system). (b)~Moving-perturbation interface (I system). (c)~Moving-perturbation periodic (bilayer unit-cell) structure (P system). (d)~Metamaterial homogenization of (c) (H system).}
\end{figure}

Figure~\ref{fig:illustration} depicts the structures of the dynamic systems under consideration, where spheres of different colors represent the atoms and molecules in the different media. In all the cases, the incidence medium, labeled as 1, is isotropic and stationary, with the refractive index of $n_1=\sqrt{\epsilon_{\r 1}\mu_{\r1}}$, where $\epsilon_{\r 1}$ and $\mu_{\r 1}$ are the relative permittivity and permeability of the first medium, respectively, while the other medium, labeled as 2, depends on the type of dynamics.

Figure~\ref{fig:illustration}(a) represents the structure of a \emph{moving-matter bloc}, referred to as ``M system''. This dynamic system involves an object made of medium 2, assumed isotropic and with $n_2=\sqrt{\epsilon_{\r 2}\mu_{\r 2}}$ at rest, propelled by an external force and moving at the velocity $\mathbf{v}$ in the background medium 1. The motion of atoms and molecules within medium 2 induces Fresnel-Fizeau drag~\cite{Fresnel_1818,Fizeau_1851}, which transforms it into a bianisotropic medium, described by $n_{2}(\mathbf{k})$~\cite{Kong_2008_emwtheory,Deck_2021_electromagnetic}.
Figure~\ref{fig:illustration}(b) shows the configuration of a \emph{moving-perturbation interface}, referred to as ``I system''. Unlike the moving-matter bloc, the interface of this system is formed by a traveling-wave modulation step. Therefore, the system does not involve matter motion, but rather the motion of the interface perturbation. This system may exceed the light speed limitation, accessing hence not only the subluminal regime but also the superluminal regime, with the limiting case  $v\rightarrow\infty$ corresponding to a pure-time interface.
Figure~\ref{fig:illustration}(c) depicts a \emph{moving-perturbation periodic structure}, referred to as ``P system''. This system is formed by a periodic modulation, composed of the bilayer unit-cell in Fig.~\ref{fig:illustration}(b). In the Bragg regime, where $l\approx\lambda_{\mathrm{i}}/2$, with $l=l_{a}+l_{b}$ representing the size of the unit cell and $\lambda_{\mathrm{i}}$ denoting the incident wavelength, the structure supports multiple scattering within each of its layer~\cite{Deck_2019_uniform}.
Finally, Fig.~\ref{fig:illustration}(d) shows an homogenized or metamaterial version of the P system, referred to as ``H system'', where the crystal's spatial unit cell is deeply sub-wavelength, i.e., $l<<\lambda_{\mathrm{i}}$. Due to its motion, this system is also bianisotropic, with effective refractive index of $n_{2\mathrm{eff}}(\mathbf{k})$~\cite{Huidobro_2021_homogenization}. 
This paper focuses on homogeneous media, including systems M, I, and H. Consequently, system P, when operating in the Bragg regime, will not be studied, as it is inhomogeneous in this regime.

\section{Generalized Energy-Momentum \\ Conservation Laws for Dynamic Systems}\label{sec:energy}
\begin{figure}[ht!]
    \centering
    \includegraphics[width=8.6cm]{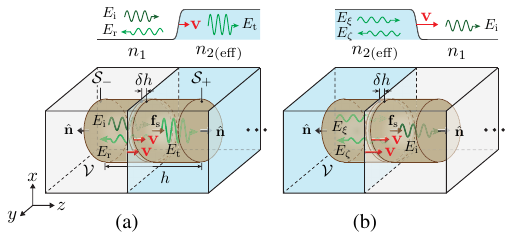}
    \caption{\label{fig:moving_cylinder} Imaginary two-media cylinder with discontinuity moving at the velocity $v$ (a)~in the subluminal regime [$v<\min(c/n_1,c/n_{2(\mathrm{eff})})$] (systems M, I and H) and (b)~in the superluminal regime [$v>\max(c/n_1,c/n_{2(\mathrm{eff})})$] (systems I and H). The top panels show the corresponding moving refractive index steps and waves.}
\end{figure}

We shall now generalize the electromagnetic energy-momentum conservation laws previously formulated for moving matter systems (M in Fig.~\ref{fig:illustration}) in~\cite{Costen_1965_three} to include both moving-perturbation interfaces (I in Fig.~\ref{fig:illustration}) and homogenized structures (H in Fig.~\ref{fig:illustration}). This will be accomplished with the help of Fig.~\ref{fig:moving_cylinder}, which depicts a two-media moving cylindrical volume, $\mathcal{V}=\mathcal{V}(z,t)$, bounded by the closed surface $\mathcal{S}=\mathcal{S}(z,t)$ with outward-pointing unit normal vector $\hat{\mathbf{n}}$. The system includes a discontinuity representing the interface between the two media, which moves with velocity $\mathbf{v}=v\hat{\mathbf{z}}$. 

The conservation law for the energy in the system of Fig.~\ref{fig:moving_cylinder} may be expressed as~\cite{Jackson_1998,Kong_2008_emwtheory}
\begin{subequations}\label{eq:p_f_moving}
    \begin{equation} \label{eq:p_moving}
    \iiint \frac{\partial W}{\partial t}~\d \mathcal{V}=-\oiint   \mathbf{S}\cdot \hat{\mathbf{n}}~\d \mathcal{S}
        +\iiint p~\d \mathcal{V},
    \end{equation}
where $W=(\D\cdot\E+\B\cdot\H)/2$ is the energy density of the wave, $\S=\E\times\H$ is the corresponding Poynting vector and $p$ is the power density gain ($p>0$) or loss ($p<0$) of the wave from the external source, which is here the source of the motion, viz., a mechanical force (system M), a step perturbation modulation (system I) or a periodic perturbation modulation (systems P and H). Equation~\eqref{eq:p_moving} means that the temporal variation of the wave energy in $\mathcal{V}$ is equal to the sum of the wave power flux through $\mathcal{S}$ into $\mathcal{V}$ and the power transferred by the external source into $\mathcal{V}$. 
    
Similarly, the conservation law for the momentum may be expressed as~\cite{Jackson_1998,Kong_2008_emwtheory}
    \begin{equation} \label{eq:f_moving}
        \iiint \frac{\partial \g}{\partial t}~\d \mathcal{V}=-\oiint \mat{\mathrm{T}}\cdot \hat{\mathbf{n}}~\d \mathcal{S}+\iiint \mathbf{f}~\d \mathcal{V},
    \end{equation}
    where $\g=\D \times \B$ is the momentum density of the wave, $\mat{\mathrm{T}}=\frac{1}{2}(\D\cdot\E+\B\cdot\H)\mat{\mathrm{I}}-\D\E-\B\H$ is the corresponding Maxwell stress tensor and $\mathbf{f}$ is the force density related to $p$ in Eq.~\eqref{eq:p_moving} by the power equation
    \begin{equation} \label{eq:p_vf}
        p=\mathbf{v}\cdot\mathbf{f},
    \end{equation}
    which states that the power density is the projection of the force onto the direction of motion multiplied by the velocity of the motion~\cite{Halliday_2013_fundamentals}.
   Equation~\eqref{eq:f_moving} means that the temporal variation of the wave momentum in $\mathcal{V}$ is equal to the sum of the wave force flux through $\mathcal{S}$ into $\mathcal{V}$ and the momentum transferred by the external source into $\mathcal{V}$. 
\end{subequations}

Equations~\eqref{eq:p_f_moving} may be rewritten, for later simplification, in terms of total (as opposed to partial) time derivatives with the help of the general Leibniz integral rule, or Reynolds transport theorem~\cite{Costen_1965_three,Landau_1984_electrodynamics,Rothwell_2018_electromagnetics},
\begin{equation}\label{eq:kinematical_theorem}
    \frac{\d}{\d t} \iiint a~\d \mathcal{V}=\iiint \frac{\partial a}{\partial t}~\d \mathcal{V}+\oiint a~\mathbf{v} \cdot \hat{\mathbf{n}}~\d \mathcal{S},
\end{equation}
where $a=a(x,y,z,t)$ may be either a scalar or vector. Using Eq.~\eqref{eq:kinematical_theorem} with $a$ replaced by $W$ and $\mathbf{g}$ in Eqs.~\eqref{eq:p_moving} and~\eqref{eq:f_moving}, respectively, yields the new conservation laws
    \begin{subequations}\label{eq:p_f_kinemic}
       \begin{equation}
        \iiint p~\d \mathcal{V} = \frac{\d}{\d t} \iiint W~\d \mathcal{V} + \oiint (\mathbf{S}-W\mathbf{v})\cdot\hat{\mathbf{n}}~\d \mathcal{S}
       \end{equation}
        and
        \begin{equation}
            \iiint \mathbf{f}~\d \mathcal{V} =\frac{\d}{\d t} \iiint  \g~\d \mathcal{V}+\oiint \left(\mat{\mathrm{T}}-\g\mathbf{v}\right)\cdot\hat{\mathbf{n}}~\d \mathcal{S}.
        \end{equation}        
    \end{subequations}

We may alternatively write the integral formulas~\eqref{eq:p_f_kinemic} into their local forms by reducing the cylinder in Fig.~\ref{fig:moving_cylinder} to an infinitesimally thin pillbox, of thickness $h=\delta{h}\rightarrow{0}$. In this case, the left-hand sides of Eqs.~\eqref{eq:p_f_kinemic} become
\begin{subequations} \label{eq:ps_fs}
    \begin{equation}
    \lim _{\delta h\rightarrow 0} \iiint p~\d \mathcal{V} =  \lim _{\delta h\rightarrow 0} \oiint \left(\int_{\delta h} p~\d h\right)\d \mathcal{S} =\lim _{\delta h\rightarrow 0} \oiint p_{\mathrm{s}}~\d \mathcal{S},
\end{equation}
and
\begin{equation}
    \lim _{\delta h\rightarrow 0} \iiint \mathbf{f}~\d \mathcal{V} =  \lim _{\delta h\rightarrow 0} \oiint \left(\int_{\delta h} \mathbf{f}~\d h\right)\d \mathcal{S} =\lim _{\delta h\rightarrow 0} \oiint \mathbf{f}_{\mathrm{s}}~\d \mathcal{S},
\end{equation}
where $p_{\mathrm{s}}=p\delta{h}$ and $\mathbf{f}_{\mathrm{s}}=\mathbf{f} \delta h$ are the \emph{surface} power and force densities corresponding to the volume power and force densities $p\rightarrow\infty$ and $\mathbf{f}\rightarrow\infty$
in the limit $\delta h\rightarrow 0$.
\end{subequations}

Substituting Eqs.~\eqref{eq:ps_fs} into Eqs.~\eqref{eq:p_f_kinemic} yields now
    \begin{subequations} \label{eq:p_f_integral}
        \begin{equation}
        \lim _{\delta h\rightarrow 0} \oiint p_{\mathrm{s}}~\d \mathcal{S} =\lim _{\delta h\rightarrow 0}\oiint (\mathbf{S}-W\mathbf{v})\cdot\hat{\mathbf{n}}~\d \mathcal{S}
        \end{equation}
        and 
        \begin{equation}
         \lim _{\delta h\rightarrow 0} \oiint \mathbf{f}_{\mathrm{s}}~\d \mathcal{S} =\lim _{\delta h\rightarrow0}\oiint \left(\mat{\mathrm{T}}-\g\mathbf{v}\right)\cdot\hat{\mathbf{n}}~\d \mathcal{S},
        \end{equation}
    \end{subequations}
where the time derivative expressions in Eqs.~\eqref{eq:p_f_kinemic} have disappeared because they are proportional to $\delta h$~\cite{Kong_2008_emwtheory}. 

Finally, using the relations
\begin{subequations} \label{eq:integral}
    \begin{equation}
    \lim _{\delta h\rightarrow 0}\oiint a~\d \mathcal{S} =\iint_{\mathcal{S}_{+}}
    a_{+}~\d \mathcal{S}+\iint_{\mathcal{S}_{-}}
    a_{-}~\d \mathcal{S}
\end{equation}
and
\begin{equation}
    \lim _{\delta h\rightarrow 0}\oiint a\cdot\hat{\mathbf{n}}~\d \mathcal{S} =\iint_{\mathcal{S}_{+}}
    a_{+}\cdot\hat{\mathbf{z}}~\d \mathcal{S}-\iint_{\mathcal{S}_{-}}
    a_{-}\cdot\hat{\mathbf{z}}~\d \mathcal{S},
\end{equation}
\end{subequations}
where $a_\pm$ represents the value of $a$ on the cylinder base surfaces\footnote{The contribution from the lateral sides vanishes because of the corresponding vanishing length $\delta{h}$.}, $\mathcal{S}_\pm$, which we assume to be equal, transforms Eqs.~\eqref{eq:p_f_integral} into their local, differential forms
\begin{subequations} \label{eq:p_f_differential}
        \begin{equation}\label{eq:power_vm}
        {\scriptstyle\sum} p_{\mathrm{s}}=\hat{\mathbf{z}} \cdot[\mathbf{S}]-v[W]
    \end{equation}
    and
        \begin{equation}\label{eq:f_vm}
         {\scriptstyle\sum} \mathbf{f}_{\mathrm{s}}=\hat{\mathbf{z}} \cdot[\mat{\mathrm{T}}]-v[\g],
    \end{equation}
    where ${\scriptstyle\sum}a= a_{+} + a_{-}$ and $[a] = a_{+} - a_{-}$ denote the sum and the difference of $a$ across the discontinuity\footnote{As seen in Eqs.~\eqref{eq:p_f_differential}, the final energy and momentum relations are actually expressed in terms of (surface) power and force densities, respectively. We will therefore often refer to power when discussing energy and to force when referring to momentum.}. ${\scriptstyle\sum}p_{\mathrm{s}}$ and ${\scriptstyle\sum}\mathbf{f}_{\mathrm{s}}$ are the sum of the surface power and force densities on the two sides of the discontinuity. Therefore, these quantities can be interpreted as the \emph{total} surface power and force densities at the discontinuity. For notational convenience, we will henceforth denote them simply as $p_{\mathrm{s}}$ and $\mathbf{f}_{\mathrm{s}}$ to represent ${\scriptstyle\sum}p_{\mathrm{s}}$ and ${\scriptstyle\sum}\mathbf{f}_{\mathrm{s}}$ from here on.
\end{subequations}

Equations~\eqref{eq:p_f_differential} are general formulas that may be interpreted in terms of wave-medium interactions as follows. In the absence of dissipation\footnote{We do not include here dispersion, which is considered negligible away from the related absorption peak.}, $p_{\mathrm{s}}$ represents the power (surface density) transferred by the external source to the wave when $p_{\mathrm{s}}>0$, or removed from the wave when $p_{\mathrm{s}}<0$. On the other hand, $\mathbf{f}_{\mathrm{s}}=f_{\mathrm{s}}\hat{\mathbf{z}}$ represents the force (surface density) exerted by the external source onto the wave when $f_{\mathrm{s}}>0$, or exerted onto the external source by the wave when $f_{\mathrm{s}}<0$.

\section{Dynamic Energy-Momentum Relations \\ under Harmonic Wave Incidence}\label{sec:harmonic}
From now on, we will restrict our attention to \emph{harmonic} waves, for which the quantities $p_{\mathrm{s}}$ and $\mathbf{f}_{\mathrm{s}}$ in Eqs.~\eqref{eq:p_f_differential} can be readily expressed in terms of the structural parameters $\eta$ (wave impedance), $n$ and $v$. 

As suggested by the insets at the top of the figures in Fig.~\ref{fig:moving_cylinder}, the scattering phenomenology differs significantly between the subluminal and superluminal\footnote{For simplicity, we will next use the terms ``reflection'' and ``transmission''. However, in the superluminal case, these should be more accurately described as ``later backward transmission'' and ``later forward transmission''~\cite{Caloz_GSTEMs,Caloz_2019_spacetime2,Deck_2019_uniform}.} cases. In both cases, incidence occurs in medium~1 and transmission occurs in medium~2. However, reflection occurs in media~1 and~2 for the subluminal and superluminal cases, respectively. The situation of the subluminal (space-like) case [Fig.~\ref{fig:moving_cylinder}(a)] is easily understandable, as it is similar to scattering at a spatial discontinuity, with reflection occurring on the same side as incidence. The superluminal (time-like) case, being similar to a temporal discontinuity with later backward and later forward waves~\cite{Caloz_2019_spacetime2}, is less intuitive. Particularly, in the comoving (incident wave
propagating in the same direction as the interface) superluminal case shown in Fig.~\ref{fig:moving_cylinder}(b), medium~1 must be placed at the \emph{right} (farther along $z$) of medium~2, for otherwise the wave, being slower than the interface, would never catch the interface and there would be no scattering, a situation that is possible but without interest; in contrast, with medium~1 on the right, the wave gets overtaken by the interface, and interesting scattering occurs~\cite{Deck_2019_uniform}. 

When the duration of the interaction, $\Delta t$, is large compared to the wave periods, we may conveniently use time averages, denoted $\langle\cdot\rangle$, in Eqs.~\eqref{eq:p_f_differential} (see Appendix~\ref{app:TV_conditions}). Substituting the time-averaged $\langle\mathbf{S}\rangle$ and $\langle W\rangle$ of the incident and scattered waves into Eq.~\eqref{eq:power_vm} and expressing all field quantities in terms of the incident electric field (see Appendix~\ref{app:p_f}), provides the time-averaged power transfer densities $\langle p_{\mathrm{s}} \rangle$ for the co/contra-moving ($\pm$) subluminal (sub) and superluminal (sup) regimes at the moving discontinuity surface as, which read
\begin{subequations} \label{eq:p_cw}
     \begin{equation} \label{eq:p_cw_sub}
       \begin{split}
        \langle p_{\mathrm{s}}^{{\mathrm{sub}\pm}} \rangle=I_{\mathrm{i}}\Bigg[\frac{\eta_1}{\eta_2}\Bigg(1-&\frac{n_{2+} v}{c}\Bigg)\tau^2 \\ &+\Bigg(1+\frac{n_{1} v}{c}\Bigg)r^2 -\Bigg(1-\frac{n_{1} v}{c}\Bigg)\Bigg]
    \end{split}  
     \end{equation}
     and
     \begin{equation} \label{eq:p_cw_sup}
        \begin{split}
        \langle p_{\mathrm{s}}^{\mathrm{sup}\pm} \rangle = \mp I_{\mathrm{i}}\Bigg\{\frac{\eta_1}{\eta_2}\Bigg[\Bigg(1-\frac{n_{2+} v}{c}\Bigg)\xi^2 - \Bigg(1+\frac{n_{2-} v}{c}\Bigg)\zeta^2\Bigg]& \\ -\Bigg(1-\frac{n_{1} v}{c}\Bigg)\Bigg\},&
        \end{split}
    \end{equation}
    where $I_{\mathrm{i}}=A_{\mathrm{i}}^2/(2 \eta_1)$ is the incident intensity, with $A_{\mathrm{i}}$ being the amplitude of the incident electric field. The exact expressions for the wave impedances of the two media, $\eta_{1,2}$, and for the refractive indices of the second medium in the $\pm z$ directions, $n_{2\pm}$, are given by Eqs.~\eqref{eq:etas} and~\eqref{eq:n2_pm}, respectively, in Appendix~\ref{app:scatt_coef}.
\end{subequations}

Similarly, substituting the time-averaged $\langle\mat{\mathrm{T}}\rangle$ and $\langle\g\rangle$ of the incident and scattered waves into Eq.~\eqref{eq:f_vm} and expressing all field quantities in terms of the incident electric field (see Appendix~\ref{app:p_f}), yields the time-averaged force densities $\langle f_{\mathrm{s}} \rangle$ at the moving discontinuity surface,
\begin{subequations} \label{eq:f_cw}
     \begin{equation} \label{eq:f_cw_sub}
       \begin{split}
        \langle f_{\mathrm{s}}^{\mathrm{sub}\pm} \rangle=\frac{I_{\mathrm{i}}}{v}\Bigg[\frac{\eta_1}{\eta_2}\Bigg(1-&\frac{n_{2+} v}{c}\Bigg)\tau^2 \\ &+\Bigg(1+\frac{n_{1} v}{c}\Bigg)r^2 -\Bigg(1-\frac{n_{1} v}{c}\Bigg)\Bigg]
    \end{split}  
     \end{equation}
     and
     \begin{equation} \label{eq:f_cw_sup}
        \begin{split}
        \langle f_{\mathrm{s}}^{\mathrm{sup}\pm} \rangle = \mp \frac{I_{\mathrm{i}}}{v}\Bigg\{\frac{\eta_1}{\eta_2}\Bigg[\Bigg(1-\frac{n_{2+} v}{c}\Bigg)\xi^2 - \Bigg(1+\frac{n_{2-} v}{c}\Bigg)\zeta^2\Bigg]& \\ -\Bigg(1-\frac{n_{1} v}{c}\Bigg)\Bigg\},&
        \end{split}
    \end{equation}
\end{subequations}
where $\eta_{1,2}$ and $n_{2\pm}$ are found as for the power case. Note that Eqs.~\eqref{eq:p_cw} and~\eqref{eq:f_cw} differ only by a factor $v$, as  expected from $\langle{p}_\mathrm{s}\rangle=\mathbf{v}\cdot\langle{\mathbf{f}}_\mathrm{s}\rangle=v\langle {f}_\mathrm{s}\rangle$ that follows from Eq.~\eqref{eq:p_vf}. These equations are valid for all the dynamic systems in Fig.~\ref{fig:illustration} (M, I and H) and will be used in the next section to investigate the wave-medium interactions in these systems.

\section{Analysis and Validation}\label{sec:analysis}
For simplicity, in this section, we focus on the impedance-matched case, where $\eta_2=\eta_1$ [Eq.~\eqref{eq:etas} in Appendix~\ref{app:scatt_coef}], leading to zero subluminal and superluminal reflection coefficients, $r=\zeta=0$ [Eqs.~\eqref{eq:r_t} in Appendix~\ref{app:scatt_coef}]. In this case, the surface power transfer densities [Eqs.~\eqref{eq:p_cw}] reduce to
\begin{subequations} \label{eq:p_cw_im}
    \begin{equation} 
        \langle p_{\mathrm{s}}^{\mathrm{sub}\pm} \rangle= I_{\mathrm{i}}\Bigg[\frac{1-n_{1}v/c}{1-(n_{1}+\Delta n)v/c}\frac{v\Delta n}{c}\Bigg]
    \end{equation}
and 
    \begin{equation} 
        \langle p_{\mathrm{s}}^{\mathrm{sup}\pm} \rangle=\mp I_{\mathrm{i}}\Bigg[\frac{1-n_{1}v/c}{1-(n_{1}+\Delta n)v/c}\frac{v\Delta n}{c}\Bigg],
    \end{equation}
\end{subequations}
and the surface force densities [Eqs.~\eqref{eq:f_cw}] reduce to
\begin{subequations} \label{eq:f_cw_im}
    \begin{equation} \label{eq:f_cw_im_sub}
        \langle f_{\mathrm{s}}^{\mathrm{sub}\pm} \rangle= I_{\mathrm{i}}\Bigg[\frac{1-n_{1}v/c}{1-(n_{1}+\Delta n)v/c}\frac{\Delta n}{c}\Bigg]
    \end{equation}
and 
    \begin{equation} \label{eq:f_cw_im_sup}
        \langle f_{\mathrm{s}}^{\mathrm{sup}\pm} \rangle=\mp I_{\mathrm{i}}\Bigg[\frac{1-n_{1}v/c}{1-(n_{1}+\Delta n)v/c}\frac{\Delta n}{c}\Bigg],
    \end{equation}
\end{subequations}
where $\Delta n = n_{2+} - n_1$ is the refractive index contrast. Given the similarity of Eqs.~\eqref{eq:p_cw_im} and~\eqref{eq:f_cw_im}, which differ only by the parameter $v$, we shall first restrict our attention to the power relations in studying wave-medium interaction in terms of $\Delta n$ (acting as an independent variable) and consider both the power and force relations in studying these interactions for the different systems (M, I and H in Fig.~\ref{fig:illustration}), where $n_{2+}$ takes various forms, generally depending on $v$ (Appendix~\ref{app:scatt_coef}).

\begin{figure}[ht!]
    \centering
    \includegraphics[width=8.6cm]{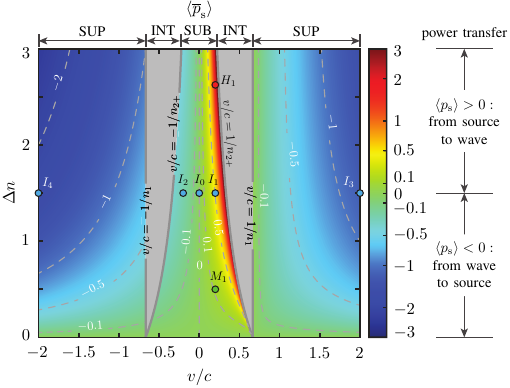}
    \caption{\label{fig:p_n2} 
    Normalized time-averaged surface power transfer density, $\langle \overline{p}_{\mathrm{s}}\rangle=\langle p_{\mathrm{s}} \rangle/I_{\mathrm{i}}$ [Eqs.~\eqref{eq:p_cw_im}], versus the normalized interface velocity, $v/c$, and the refractive index contrast, $\Delta n=n_{2+}-n_1$, with $n_1=1.5$ and $\eta_2=\eta_1$ (impedance-matching) for all the systems (M, I and H) in Fig.~\ref{fig:illustration}. The gray areas correspond to the interluminal regimes $|c/n_{2+}|<|v|<|c/n_1|$. The shorthand notations ``SUB'', ``INT'' and ``SUP'' at the top of the graph refer to the subluminal, interluminal and superluminal regimes, respectively. The colored points labeled $I_0,I_1,I_2,I_3,I_4,M_1$ and $H_1$ will be discussed in connection with Figs.~\ref{fig:catch_up} and~\ref{fig:FF_drag}.}
\end{figure}

Equations~\eqref{eq:p_cw_im} indicate that, if medium 1 is fixed, the surface power transfer density depends only on $\Delta n$ and $v$. Figure~\ref{fig:p_n2} plots the corresponding normalized surface power transfer density, $\langle \overline{p}_{\mathrm{s}}\rangle=\langle p_{\mathrm{s}}\rangle/I_{\mathrm{i}}$, given by Eqs.~\eqref{eq:p_cw_im}, versus the normalized interface velocity, $v/c$, and the refractive index contrast\footnote{We assume that $\Delta n>0$ throughout the paper.}, $\Delta n$, for the impedance-matched case.

\begin{figure}[ht!]
    \centering
    \includegraphics[width=8.6cm]{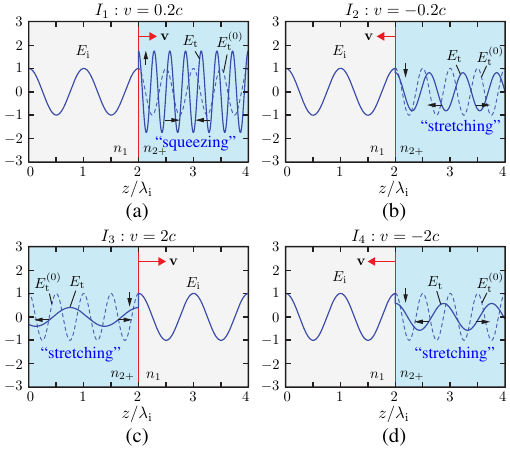}
    \caption{\label{fig:catch_up} Illustration of the ``pushing-squeezing'' and ``pulling-stretching'' effects of the moving interface on the transmitted wave for the (a)~comoving subluminal case, $v=0.2c$ (corresponding to point $I_1$ in Fig.~\ref{fig:p_n2}), (b)~contramoving subluminal case, $v=-0.2c$ (corresponding to point $I_2$ in Fig.~\ref{fig:p_n2}), (c)~comoving superluminal case, $v=2c$ (corresponding to point $I_3$ in Fig.~\ref{fig:p_n2}) and (d)~contramoving superluminal case, $v=-2c$ (corresponding to point $I_4$ in Fig.~\ref{fig:p_n2}). The dashed curves in the second medium represent the transmission in the stationary case ($v=0$, corresponding to point $I_0$ in Fig.~\ref{fig:p_n2}) for comparison.}
\end{figure}

Figure~\ref{fig:p_n2} shows that, for a given $\Delta n$, the sign--and consequently the direction--of the power transfer varies across four distinct  velocity regimes, namely, the comoving/contramoving and subluminal/superluminal regimes. This variation arises from ``pushing'' and ``pulling'' effects\footnote{A similarly intuitive description of space-time interaction effects was presented in~\cite{Deck_2019_uniform,Li_2023_TIR}: the ``catch-up'' effect, whereby the incident wave catches up to a slower modulation interface, leading to scattering interaction, but does not catch up to a faster modulation interface, resulting in no interaction.} of the moving interface onto the transmitted wave in the different velocity regimes, as illustrated in Fig.~\ref{fig:catch_up}.
In the comoving ($v>0$) subluminal regime, the moving interface ``pushes'' the transmitted wave, as shown in Fig.~\ref{fig:catch_up}(a), resulting in an increase of both frequency and amplitude (wave squeezing or compression in~\cite{Caloz_2019_spacetime2}), with power transferred from the external source to the wave ($p_{\mathrm{s}}>0$). In the other velocity regimes, the moving interface ``pulls'' the transmitted wave, as shown in Figs.~\ref{fig:catch_up}(b)-(d), resulting in a decrease of both frequency and amplitude (wave stretching or expansion in~\cite{Caloz_2019_spacetime2}), with power transferred from the wave to the external source ($p_{\mathrm{s}}< 0$). In each velocity regime, the effect intensifies as the interface velocity increases, leading to greater power gain or loss, as could be expected.

On the other hand, at a given velocity, $v$, the magnitude of $\langle \overline{p}_{\mathrm{s}}\rangle$ increases as $\Delta n$ increases in all the velocity regimes. This is because the variation of $\Delta n$ influences the transmitted wave velocity, $v_{2+}=c/n_{2+}$, which in turn affects the velocity difference between the moving interface (fixed $v$) and the transmitted wave, given by $\Delta v=|v-v_{2+}|$. These changes ultimately alter the ``pushing'' and ``pulling'' effects of the moving interface onto the transmitted wave, leading to variations in power gain and loss.
For instance, in the comoving subluminal regime, an increase in $\Delta n$ results in a decrease in $v_{2+}$ and a reduction\footnote{In the comoving subluminal regime shown in Fig.~\ref{fig:p_n2}, where $0<v<v_{2+}$, we have $\Delta v=|v-v_{2+}|=v_{2+}-v$.} in $\Delta v$. In this scenario, the ``pushing'' effect increases, leading to higher transmitted frequency and amplitude, ultimately resulting in more significant power gain. 
This effect intensifies with increasing $\Delta n$ in each velocity regime, leading to greater power gain or loss.

\begin{figure}[ht!]
    \centering
    \includegraphics[width=8.6cm]{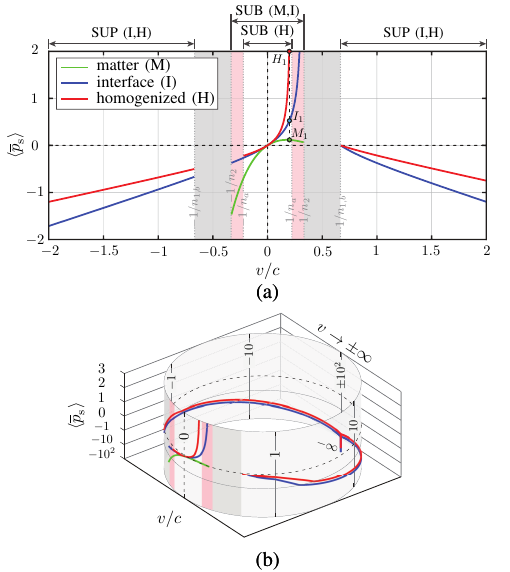}
    \caption{\label{fig:p_plot} Normalized time-averaged surface power transfer densities, $\langle \overline{p}_{\mathrm{s}}\rangle=\langle p_{\mathrm{s}} \rangle/I_{\mathrm{i}}$, versus the normalized interface velocity, $v/c$, for the dynamic systems in Fig.~\ref{fig:illustration} (a)~in a flat graph and (b)~on a cylindrical surface emphasizing the convergence of the comoving and contramoving superluminal regimes to the same state at $v\rightarrow\pm\infty$. The systems M and I have the rest medium parameters $\epsilon_{\mathrm{r}1}=\mu_{\mathrm{r}1}=1.5$ and  $\epsilon_{\mathrm{r}2}=\mu_{\mathrm{r}2}=3$, and the H system has the rest medium parameters $\epsilon_{\mathrm{r}1,b}=\mu_{\mathrm{r}1,b}=1.5$ and $\epsilon_{\mathrm{r}a}=\mu_{\mathrm{r}a}=4.5$, with $l_a=l_b$ [Fig.~\ref{fig:illustration}(c)]. A log scale is used in Fig.~\ref{fig:p_plot}(b) for $\langle \overline{p}_{\mathrm{s}} \rangle<-1$ and $|v/c|>1$ to show the limiting case of $v\rightarrow\pm\infty$. The colored points labeled $I_1,M_1$ and $H_1$ in Fig.~\ref{fig:p_plot}(a) will be discussed in connection with Figs.~\ref{fig:p_n2} and~\ref{fig:FF_drag}.}
\end{figure}

Figure~\ref{fig:p_plot} compares the normalized surface power transfer densities [Eqs.~\eqref{eq:p_cw_im} with $n_{2+}$ given by Eqs.~\eqref{eq:n2_pm} in Appendix~\ref{app:scatt_coef}] for the three systems, M, I and H, in different velocity regimes\footnote{In the case of the H system, we simplify the analysis by assuming that medium $b$ is identical to medium 1 [Fig.~\ref{fig:illustration}(c)], resulting in $\epsilon_b=\epsilon_1$ and $\mu_b=\mu_1$. For the sake of comparison, we set $\epsilon_a=2\epsilon_2-\epsilon_1$ and $\mu_a=2\mu_2-\mu_1$ to ensure the effective refractive index at rest, $n_{2+}(v=0)=c\sqrt{\overline{\epsilon}_{\|}\overline{\mu}_{\|}}$, with $\overline{\epsilon}_{\|}=(\epsilon_a+\epsilon_b)/2$, $\overline{\mu}_{\|}=(\mu_a+\mu_b)/2$ [Eqs.~\eqref{eq:para_K_sub} and~\eqref{eq:para_Kp_sub}]~\cite{Huidobro_2019_fresnel,Bahrami_2023_electrodynamics}, are identical to that of M and I systems, $n_{2+}(v=0)=c\sqrt{\epsilon_{2}\mu_{2}}$ [Eqs.~\eqref{eq:para_matter} and~\eqref{eq:para_interface}].}, still assuming impedance matching. The dashed areas in the figure denote the interluminal regimes\footnote{Note that the interluminal regimes differ among systems. In the case of the system H, where three media are involved before homogenization [Fig.~\ref{fig:illustration}(c)], to ensure effective homogenization~\cite{Huidobro_2021_homogenization}, the interluminal velocity regime is characterized by $|\min(c/n_1,c/n_a,c/n_b)|<|v|<|\max(c/n_1,c/n_a,c/n_b)|$, where $n_a=\sqrt{\epsilon_{\r a}\mu_{\r a}}$ and $n_b=\sqrt{\epsilon_{\r b}\mu_{\r b}}$ are the refractive indices of the periodic modulated media in Fig.~\ref{fig:illustration}(c). In contrast, for the systems involving only two media, such as systems M and I [Figs.~\ref{fig:illustration}(a) and~(b)], the velocity regime is $|\min(c/n_1,c/n_2)|<|v|<|\max(c/n_1,c/n_2)|$.}, encompassing phenomena such as \v{C}erenkov radiation (system M)~\cite{Kong_2008_emwtheory,Ostrovskii_1975_some} and other less understood physics (systems I and H)~\cite{Deck_2019_uniform,Huidobro_2021_homogenization}, which is beyond the scope of this paper.

Several comments are in order concerning Fig.~\ref{fig:p_plot}. A global observation is that the moving-perturbation systems (I and H) achieve power transfer similar to the moving-matter system (M) at ``small'' velocities of up to about $10\%$ of $c$. In contrast, at large velocities in the superluminal regime ($v\rightarrow\pm\infty$), the powers tend to negative infinity, theoretically suggesting infinite power transfer from the wave to the modulation~\cite{Tan_2020_energy}. However, this does \emph{not} mean that energy transfer is infinite, since the energy $\Delta\mathcal{E}=p_\mathrm{s}\Delta{t}$ with $\Delta{t}\rightarrow{0}$ allows $\Delta\mathcal{E}$ to be finite even for $p_\mathrm{s}\rightarrow\infty$\footnote{For instance, the purely temporal ($v=\infty$) transmission coefficient for the system I under impedance matching is $\xi=n_1/n_2=1.5/3=0.5$ [Eq.~\eqref{eq:sup_r_t}], which indicates \emph{finite} energy transfer.}.

We further observe that the surface power transfer density varies among the different dynamic systems at a given velocity, $v$, as shown in Fig.~\ref{fig:p_plot}. This variation is due to differences in the refractive index contrasts, $\Delta n$, caused by the Fresnel-Fizeau drag in these systems.
Figure~\ref{fig:FF_drag} presents the refractive index of medium 2, $n_{2+}$ [Eq.~\eqref{eq:n2_pm} in Appendix~\ref{app:scatt_coef}], versus the normalized interface velocity, $v/c$, for the three considered systems. 
For instance, at the comoving subluminal velocity $v=0.2c$, whose isofrequencies are plotted in the inset of Fig.~\ref{fig:FF_drag}, the conventional Fresnel-Fizeau drag in the M system leads to a smaller $n_{2+}$ compared to that of the I system, which experiences no drag. In contrast, the drag due to the periodic modulation, acting in the opposite direction to the motion in the H system, results in an increase of $n_{2+}$ [see the corresponding set of points $(M_1,I_1,H_1)$ marked on the curves at $v=0.2c$ in Fig.~\ref{fig:FF_drag}]. 
As discussed in Eqs.~\eqref{eq:p_cw_im} and Fig.~\ref{fig:p_n2}, the variations in $n_{2+}$, leading to changes in $\Delta n$, result in differing power transfers among the various systems. Specifically, largest and smallest power transfer occur the H and M systems, respectively, as indicated by the corresponding points $(M_1,I_1,H_1)$ in Figs.~\ref{fig:p_n2} and~\ref{fig:p_plot}(a).
Table~\ref{tab:analysis} summarizes these results.

\begin{figure}[ht!]
    \centering
    \includegraphics[width=8.6cm]{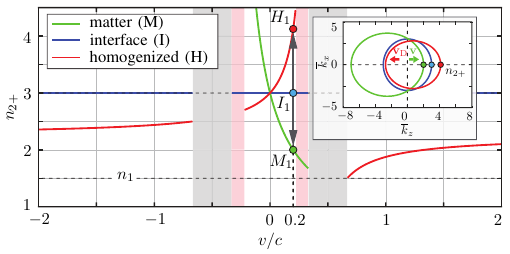}
    \caption{\label{fig:FF_drag} Refractive index of medium 2, $n_{2+}$ [Eq.~\eqref{eq:n2_pm} in Appendix~\ref{app:scatt_coef}], versus the normalized interface velocity, $v/c$, for the dynamic systems in Fig.~\ref{fig:illustration}. The inset shows the isofrequency curves in the $k_x-k_z$ plane for $v=0.2c$, along with the corresponding conventional Fresnel-Fizeau drag induced by moving matter with velocity $\mathbf{v}$ in the M system (green arrow)~\cite{Kong_2008_emwtheory,Deck_2021_electromagnetic} and the modulation-induced drag in the H system with the opposite equivalent ``moving-matter'' velocity $\mathbf{v}_{\mathrm{D}}$ (red arrow)~\cite{Huidobro_2019_fresnel}.}
\end{figure}
\begin{table}[ht!]
\centering
\caption{Comparison of the three systems at a fixed velocity in the comoving subluminal regime with arrows indicating differences with respect to the I (reference) case.}
\label{tab:analysis}
\setlength{\tabcolsep}{3pt}
\begin{NiceTabular}{c||c|c|c}
\begin{minipage}[c]{.42\columnwidth}
\centering
\vspace{3pt}
    fixed $n_1$ and $v$ 
\vspace{3pt}
\end{minipage} & \begin{minipage}[c]{.17\columnwidth}
\centering
\vspace{3pt}
    M 
\vspace{3pt}
\end{minipage} & \begin{minipage}[c]{.1\columnwidth}
\centering
\vspace{3pt}
    I
\vspace{3pt}
\end{minipage} & \begin{minipage}[c]{.17\columnwidth}
\centering
\vspace{3pt}
    H 
\vspace{3pt}
\end{minipage} \\ 
\hline\hline
\begin{minipage}[c]{.42\columnwidth}
\centering
\vspace{2pt}
    Fresnel-Fizeau drag
\vspace{2pt}
\end{minipage}
& 
$>0$~\cite{Fresnel_1818,Fizeau_1851} & $0$ & $<0$~\cite{Huidobro_2019_fresnel,Caloz_GSTEMs}
\\ \hline
\begin{minipage}[c]{.42\columnwidth}
\centering
\vspace{2pt}
    wave velocity in the \\ $+z$-direction, $v_{2+}$
\vspace{2pt}
\end{minipage} & 
$\uparrow$ & $-$ & $\downarrow$ 
\\ \hline
\begin{minipage}[c]{.42\columnwidth}
\centering
\vspace{2pt}
    refractive index, \\ $n_{2+}=c/v_{2+}$
\vspace{2pt}
\end{minipage} & 
$\downarrow$ & $-$ & $\uparrow$ 
\\ \hline
\begin{minipage}[c]{.42\columnwidth}
\centering
\vspace{2pt}
    refractive index contrast, \\ $\Delta n=n_{2+}-n_1$
\vspace{2pt}
\end{minipage}
& 
$\downarrow$ & $-$ & $\uparrow$ 
\\ \hline
\begin{minipage}[c]{.42\columnwidth}
\centering
\vspace{2pt}
    surface power and force density, \\ $\langle p_{\mathrm{s}} \rangle$, $\langle f_{\mathrm{s}} \rangle$
\vspace{2pt}
\end{minipage}
 & 
$\downarrow$ & $-$ & $\uparrow$ \\[1pt]
\end{NiceTabular}
\end{table}

\begin{figure}[ht!]
    \centering
    \includegraphics[width=8.6cm]{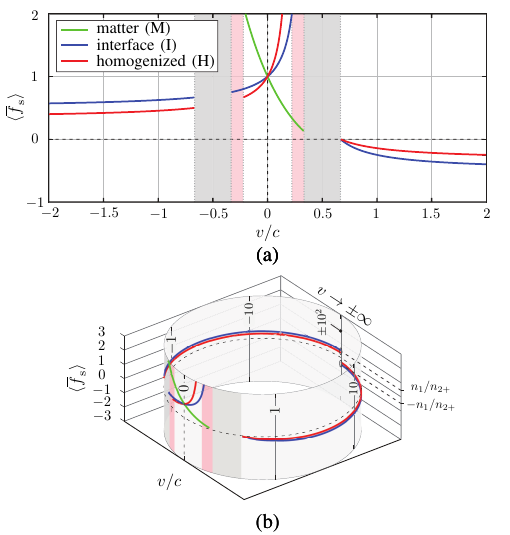}
    \caption{\label{fig:f_plot} Normalized time-averaged surface force densities, $\langle \overline{f}_{\mathrm{s}} \rangle=\langle f_{\mathrm{s}} \rangle/ \langle f_{0} \rangle$ [Eqs.~\eqref{eq:f_cw_im}], with $\langle f_{0} \rangle$ being the surface force density at $v=0$, versus the normalized interface velocity, $v/c$, for the different dynamic systems in Fig.~\ref{fig:illustration} (a)~in a flat graph and (b)~on a cylindrical surface emphasizing the opposite momenta of the comoving and contramoving superluminal regimes at $v\rightarrow\pm{\infty}$. The media parameters are identical to those used in Fig.~\ref{fig:p_plot}. A log scale is used in Fig.~\ref{fig:f_plot}(b) for $|v/c|>1$.}
\end{figure}

Figure~\ref{fig:f_plot} compares the surface force densities, $\langle f_{\mathrm{s}} \rangle$ [Eqs.~\eqref{eq:f_cw_im}], normalized by the force density at $v=0$, $\langle f_{0} \rangle$, for the three systems in different velocity regimes. The media parameters are identical to those used in Fig.~\ref{fig:p_plot}.
As shown in the figure, in the pure-space case, where $v=0$, the surface force density, given by $\langle f_{0} \rangle=I_{\mathrm{i}}\Delta n/c$ [Eq.~\eqref{eq:f_cw_im_sub}], is nonzero. This is due to the spatial variation of the media, implying momentum non-conservation~\cite{Noether_1918_invariante,Caloz_2019_spacetime2}. Furthermore, it follows from $\Delta n>0$ that $\langle f_{0} \rangle>0$.
This is due to the increase in wavenumber after the wave interacts with the interface, specifically that the transmitted wavenumber, $k_{\t}$, is larger than the incident wavenumber, $k_{\mathrm{i}}$ (see Fig.~\ref{fig:catch_up}), which results in an increase in wave momentum, i.e., $\langle f_{0} \rangle>0$.
Moreover, as shown in Fig.~\ref{fig:f_plot}(b), in the limiting case of $v\rightarrow\pm\infty$, the corresponding forces, $\langle \overline{f}_{\pm\infty} \rangle\rightarrow\mp n_1/n_{2+}$ [Eq.~\eqref{eq:f_cw_im_sup}], are equal in magnitude but opposite in sign, resulting in a net force of zero at $v=\pm\infty$, consistently with the expected conservation of momentum in the pure-time case, due to spatial translational symmetry~\cite{Noether_1918_invariante,Francesco_2022_energy,Ortega_2023_tutorial}.

Figure~\ref{fig:f_plot}(a) further reveals that the surface force density varies among the different dynamic systems at a given velocity, $v$. This variation arises from drag-induced differences from the refractive index, $n_{2+}$ (see Fig.~\ref{fig:FF_drag}), resulting wavenumber variations and, consequently, force density changes.
In the comoving subluminal regime, the transmitted wavenumber, $k_{\t}=\omega_{\t}n_{2+}/c$, increases for the system I due to the increase in transmitted frequency [see Fig.~\ref{fig:catch_up}(a)], resulting in an increase of $f_{\mathrm{s}}$. A greater increase in $f_{\mathrm{s}}$ occurs in the system H due to the greater increase of $n_{2+}$, caused by the modulation-induced drag (see the inset of Fig.~\ref{fig:FF_drag}). In contrast, the refractive index, $n_{2+}$, of the system M decreases due to the conventional drag (see the inset of Fig.~\ref{fig:FF_drag}), leading to a decrease in $k_{\t}$ and $f_{\mathrm{s}}$. 
These effects are reversed in the contramoving subluminal regime, where the drag effects act in the opposite direction. 
In the superluminal regime, the wave experiences the ``pulling'' effect represented in Figs.~\ref{fig:catch_up}(c) and (d), and resulting in a decrease in wavenumber and $f_{\mathrm{s}}$.

\begin{figure}[ht!]
    \centering
    \includegraphics[width=8.6cm]{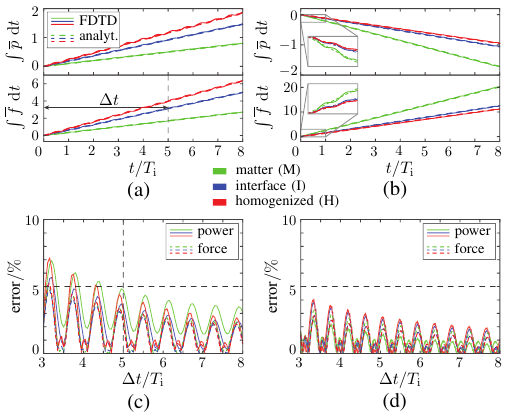}
    \caption{\label{fig:FDTD_sub} Full-wave (FDTD) simulations of the normalized time-integrated power transfer and force densities versus time for the different dynamic systems with (a)~$v=0.1c$ and (b)~$v=-0.1c$, with panels (c) and (d) showing the corresponding error versus the normalized interaction duration, $\Delta t/T_{\mathrm{i}}$, where $T_{\mathrm{i}}$ is the incident period. The media parameters are identical to those used in Fig.~\ref{fig:p_plot}.}
\end{figure}

Figure~\ref{fig:FDTD_sub} presents the full-wave simulations for the three systems with $v=0.1c$ and $v=-0.1c$, using the modified FDTD scheme recently proposed in~\cite{Deck_2023_yeecell,Bahrami_2023_FDTD}, to validate the accuracy of the time averages used in the derivations of power and force densities (Sec.~\ref{sec:harmonic} and Appendix~\ref{app:TV_conditions}).
Figures~\ref{fig:FDTD_sub}(a) and (b) provide the power transfer (top panels) and force (bottom panels) densities integrated over time--representing the total energy and momentum transfers--for the different systems. These figures compare the simulation results obtained using the original conservation laws [Eqs.~\eqref{eq:p_moving} and~\eqref{eq:f_moving}] with the analytical time-averaged formulas~\eqref{eq:p_cw} and~\eqref{eq:f_cw}.
As shown in the insets of Fig.~\ref{fig:FDTD_sub}(b), the analytical curves correspond to the averages of the related oscillating simulation curves.
To assess the accuracy of the time-averaged approximation, Figs.~\ref{fig:FDTD_sub}(c) and (d) plot the corresponding error versus the interaction duration, $\Delta t$. The error is defined as the absolute value of the difference between the analytical results, $\langle p_{\mathrm{s}}\rangle$ given in Eqs.~\eqref{eq:p_cw} and $\langle f_{\mathrm{s}}\rangle$ given in Eqs.~\eqref{eq:f_cw}, and the simulation results averaged over $\Delta t$, given by $\int p~\d t/\Delta t$ and $\int f~\d t/\Delta t$, normalized to the corresponding analytical results.
As shown in the figure, the error curves exhibit an oscillatory decrease versus $\Delta t$, and go below $5\%$ at around $\Delta t>5T_{\mathrm{i}}$, confirming the accuracy of the time-averaged approximations used in the derivations of the power transfer and force densities [Eqs.~\eqref{eq:p_cw} and~\eqref{eq:f_cw}] at large time scales.

%

\section{Conclusion}\label{sec:conclusion}
This work has filled a gap in understanding wave-medium interactions in dynamic systems. 
Deriving general solutions and analyzing energy-momentum relations, we have found that the interface velocity and the refractive index contrast of the media are two important factors influencing interactions. Specifically, comparing the different dynamic systems, we discovered that the Fresnel-Fizeau drag accounts for the variations in interactions across the systems.
These may greatly help development practical implementations of \emph{modulation} systems.

\appendices
\section{Time Averages for Harmonic Dynamic Systems}\label{app:TV_conditions}

In \emph{stationary} harmonic systems, time averages can be employed when the observation time scale significantly exceeds the wave's period. This approach simplifies the analysis of energy and momentum relations by averaging out rapid oscillations~\cite{Born_1999_principles}. However, in \emph{dynamic} harmonic systems, where the wave frequency is not conserved, time-averaging becomes more challenging. Nonetheless, we will demonstrate here that it remains feasible under certain specific conditions.

Assume the time-harmonic electric field function
\begin{equation} \label{eq:harmonic_E}
    E_{\mathrm{i,r,t}}=A_{\mathrm{i,r,t}}\cos(\omega_{\mathrm{i,r,t}}t+\phi_{\mathrm{i,r,t}}),
\end{equation}
where $A_{\mathrm{i,r,t}}$, $\omega_{\mathrm{i,r,t}}$ and $\phi_{\mathrm{i,r,t}}$ are the amplitude, frequency and initial phase of the incident, reflected and transmitted fields, respectively. Next consider, for instance, the subluminal regime in the general formula~\eqref{eq:power_vm}. Substituting Eq.~\eqref{eq:harmonic_E} into Eq.~\eqref{eq:power_vm} under the subluminal assumption, and averaging the resulting time integral over the interaction duration, $\Delta t=t_2-t_1$, yields the exact time-averaged power transfer density
\begin{equation} \label{eq:ps_ta_exact}
\begin{split}
    \langle p_{\mathrm{s}} \rangle=&\frac{A_{\t}^2}{\eta_2}\Bigg(1-\frac{n_{2+}v}{c}\Bigg)\frac{\int_{t_1}^{t_2}\cos^2(\omega_{\t}t+\phi_{\t})~\d t}{\Delta t}\\ 
    &+\frac{A_{\r}^2}{\eta_1}\Bigg(1+\frac{n_{1}v}{c}\Bigg)\frac{\int_{t_1}^{t_2}\cos^2(\omega_{\r}t+\phi_{\r})~\d t}{\Delta t}\\
    &-\frac{A_{\mathrm{i}}^2}{\eta_1}\Bigg(1-\frac{n_{1}v}{c}\Bigg)\frac{\int_{t_1}^{t_2}\cos^2(\omega_{\mathrm{i}}t+\phi_{\mathrm{i}})~\d t}{\Delta t},
\end{split}
\end{equation}
where the relations between the wave amplitudes $A_{\mathrm{i,r,t}}$ and the exact expression for the refractive index of the second medium in the $+ z$ direction, $n_{2+}$, are given by Eqs.~\eqref{eq:r_t} and~\eqref{eq:n2_pm} in Appendix~\ref{app:scatt_coef}. Note that all the expressions outside of the integrands in Eq.~\eqref{eq:ps_ta_exact} are time independent.

Applying the trigonometric identity $\cos^2\theta=(1+\cos2\theta)/2$ to the time-averaged integral terms in Eq.~\eqref{eq:ps_ta_exact} and solving the resulting integrals with the relation $T_{\mathrm{i},\r,\t}=2\pi/\omega_{\mathrm{i},\r,\t}$, we find
\begin{equation} \label{eq:ta_cos}
\begin{split}
    \frac{1}{\Delta t}\int_{t_1}^{t_2}&\cos^2(\omega_{\mathrm{i},\r,\t}t+\phi_{\mathrm{i},\r,\t})~\d t\\
    &=\frac{1}{2}+\frac{1}{2\Delta t}\int_{t_1}^{t_2}\cos(2\omega_{\mathrm{i},\r,\t}t+2\phi_{\mathrm{i},\r,\t})~\d t\\
    &=\frac{1}{2}+\frac{\sin(2\omega_{\mathrm{i},\r,\t}t+2\phi_{\mathrm{i},\r,\t})|_{t=t_1}^{t=t_2}}{8\pi}\frac{T_{\mathrm{i},\r,\t}}{\Delta t}.
\end{split}
\end{equation}

If $\Delta t$ is much larger than the wave periods $T_{\mathrm{i},\r,\t}$, viz., $\Delta t\gg T_{\mathrm{i},\r,\t}$, the second term on the right-hand side of the last equality in Eq.~\eqref{eq:ta_cos} is negligible compared to $1/2$.
Equation~\eqref{eq:ta_cos} reduces then to
\begin{equation} \label{eq:ta_half}
    \frac{1}{\Delta t}\int_{t_1}^{t_2}\cos^2(\omega_{\mathrm{i},\r,\t}t+\phi_{\mathrm{i},\r,\t})~\d t\overset{\Delta t\gg T_{\mathrm{i},\r,\t}}{\sim}\frac{1}{2}.
\end{equation}
Substituting Eq.~\eqref{eq:ta_half} into Eq.~\eqref{eq:ps_ta_exact} yields Eqs.~\eqref{eq:p_cw_sub}. Thus, time averaging remains valid in dynamic harmonic systems, with changed wave periods, under the condition of large interaction duration versus the largest wave period. The same procedure and conclusion apply to the other velocity regimes.

\section{Scattering at Dynamic Discontinuities \\ and Media Properties}\label{app:scatt_coef}

The scattering problem at the dynamic discontinuities shown in Fig.~\ref{fig:illustration} can be conceptualized as a general scattering problem of a moving discontinuity separating an isotropic medium (medium 1, with $\epsilon_1$ and $\mu_1$) and a bianisotropic medium (medium 2, with the $3\times3$ constitutive tensors $\mat{\epsilon}_2,\mat{\mu}_2,\mat{\xi}_2$ and $\mat{\zeta}_2$ being the permittivity, permeability, magnetic-to-electric and electric-to-magnetic cross-coupling tensors~\cite{Kong_2008_emwtheory,Caloz_2019_spacetime1,Deck_2021_electromagnetic}).

\subsection{Media Parameters}

We provide here the medium parameters of medium~2 for the three dynamic systems (M, I and H) in Fig.~\ref{fig:illustration}.

The matter-motion-induced bianisotropic tensors for the system M~[Fig.~\ref{fig:illustration}(a)] are~\cite{Deck_2021_electromagnetic}
\begin{subequations} \label{eq:para_all}
\begin{equation} \label{eq:para_matter}
    \begin{gathered}
        \mat{\epsilon}_2=\epsilon_2 \left[\begin{array}{lll}
        \alpha_2 & 0 & 0 \\
        0 & \alpha_2 & 0 \\
        0 & 0 & 1
    \end{array}\right],\quad 
        \mat{\mu}_2=\mu_{2}\left[\begin{array}{lll}
        \alpha_2 & 0 & 0 \\
        0 & \alpha_2 & 0 \\
        0 & 0 & 1
    \end{array}\right],
    \\
        \mat{\xi}_2=\left[\begin{array}{ccc}
        0 & \xi_2 & 0 \\
        -\xi_2 & 0 & 0 \\
        0 & 0 & 0
    \end{array}\right] \quad\text{and}\quad \mat{\zeta}_2=-\mat{\xi}_2,
    \end{gathered}
\end{equation}
where $\alpha_2=\frac{1-(v/c)^2}{1-( n_2 v/c)^2}$ and $\xi_2=\frac{v}{c^2}\frac{1-n_{2}^{2}}{1-( n_2 v/c)^2}$.

In the system I [Fig.~\ref{fig:illustration}(b)], the second medium is stationary and isotropic~\cite{Deck_2019_uniform}, so that
    \begin{equation} \label{eq:para_interface}    \mat{\epsilon}_2=\epsilon_2\mat{\mathrm{I}},\quad\mat{\mu}_2=\mu_2\mat{\mathrm{I}}\quad \textrm{and} \quad\mat{\xi}_2=\mat{\zeta}_2=\mat{\mathrm{O}},
    \end{equation}
where $\mat{\mathrm{I}}$ and $\mat{\mathrm{O}}$ are the $3\times3$ identity and null tensors, respectively.

The periodic-modulation-induced bianisotropic tensors for the system H [Fig.~\ref{fig:illustration}(d)] in the sub-wavelength regime are given as
\begin{equation} \label{eq:para_homo}
    \begin{gathered}
        \mat{\epsilon}_2=\left[\begin{array}{lll}
        \overline{\epsilon}_{\|} & 0 & 0 \\
        0 & \overline{\epsilon}_{\|} & 0 \\
        0 & 0 & \overline{\epsilon}_{\perp}
        \end{array}\right],\quad 
        \mat{\mu}_2=\left[\begin{array}{lll}
        \overline{\mu}_{\|} & 0 & 0 \\
        0 & \overline{\mu}_{\|} & 0 \\
        0 & 0 & \overline{\mu}_{\perp}
        \end{array}\right],
        \\
        \mat{\xi}_2=\left[\begin{array}{ccc}
        0 & \overline{\xi} & 0 \\
        -\overline{\xi} & 0 & 0 \\
        0 & 0 & 0
        \end{array}\right] \quad\text{and}\quad \mat{\zeta}_2=-\mat{\xi}_2,
    \end{gathered}
\end{equation}
with the homogenized medium parameters~\cite{Huidobro_2021_homogenization,Bahrami_2023_electrodynamics}
\begin{equation} \label{eq:para_K_sub}
    \begin{gathered}
    \overline{\epsilon}_{\|} =\overline{\epsilon}'_{\|}\frac{1-(v/c)^2}{\left(1+\overline\xi' v \right)^2-\overline{\epsilon}'_{\|} \overline{\mu}'_{\|} v^2}, \\
    \overline{\mu}_{\|}=\overline{\mu}'_{\|} \frac{1-(v/c)^2}{\left(1+\overline\xi' v \right)^2-\overline{\epsilon}'_{\|} \overline{\mu}'_{\|} v^2}, \\
    \overline{\xi}=\frac{\left(\overline\xi'+v/c^2\right)\left(1+\overline\xi' v\right)-\overline\epsilon'_{\|} \overline\mu'_{\|} v}{\left(1+\overline\xi' v\right)^2-\overline\epsilon'_{\|} \overline\mu'_{\|} v^2}, \\
    \overline{\epsilon}_{\perp} = \left(\frac{f}{\epsilon_{a}} +\frac{1-f}{\epsilon_{b}}\right)^{-1} \quad \textrm{and} \quad \overline{\mu}_{\perp} = \left(\frac{f}{\mu_{a}} +\frac{1-f}{\mu_{b}}\right)^{-1},
    \end{gathered}
\end{equation}
with 
\begin{equation} \label{eq:para_Kp_sub}
    \begin{gathered}
        f=l_a/(l_a+l_b),\\
        \overline{\epsilon}'_{\|}=f\frac{1-(v/c)^2}{1-( n_a v/c)^2}\epsilon_{a}+(1-f)\frac{1-(v/c)^2}{1-( n_b v/c)^2}\epsilon_{b}, \\
        \overline{\mu}'_{\|}=f\frac{1-(v/c)^2}{1-( n_a v/c)^2}\mu_{a}+(1-f)\frac{1-(v/c)^2}{1-( n_b v/c)^2}\mu_{b}, \\
        \overline{\xi}'=-f\frac{v}{c^2}\frac{1-n_{a}^{2}}{1-( n_a v/c)^2}-(1-f)\frac{v}{c^2}\frac{1-n_{b}^{2}}{1-( n_b v/c)^2}.
     \end{gathered}
\end{equation}
\end{subequations}

\subsection{General Formulas for Frequency Transitions and Scattering Coefficients at Dynamic Discontinuities}

The frequency relations between the scattered waves and incident wave are derived by transforming the laboratory-frame ($K$) fields to the comoving frame ($K'$) in the subluminal regime, where the interfaces are purely spatial and the frequencies are conserved ($\omega'_{\r}=\omega'_{\t}=\omega'_{\mathrm{i}}$), and to the instantaneous frame ($K'$) in the superluminal regime, where the interfaces are purely temporal and the wavenumbers are conserved ($k'_{\zeta}=k'_{\xi}=k'_{\mathrm{i}}$)~\cite{Deck_2019_uniform}. Then, transforming the conserved values back to the $K$ frame using Lorentz transformations $\omega^{\pm\prime}=\gamma(\omega^{\pm}\mp v_{\mathrm{f}} k^{\pm})$ and $k^{\pm\prime}=\gamma(k^{\pm}\mp v_{\mathrm{f}}\omega^{\pm}/c^2)$~\cite{Deck_2019_uniform,Kong_2008_emwtheory} gives the subluminal frequency transition relations
\begin{subequations} \label{eq:w}
    \begin{equation} \label{eq:sub_w}
    \frac{\omega_{\r}}{\omega_{\mathrm{i}}}=\frac{1-n_1 v/c}{1+n_1 v/c} 
    \quad \textrm{and} \quad 
    \frac{\omega_{\t}}{\omega_{\mathrm{i}}}=\frac{1-n_1 v/c}{1-n_{2+} v/c},
\end{equation}
and the superluminal relations
\begin{equation} \label{eq:sup_w}
    \frac{\omega_{\zeta}}{\omega_{\mathrm{i}}}=\frac{1-n_1 v/c}{1+n_{2-} v/c}
    \quad \textrm{and} \quad
    \frac{\omega_{\xi}}{\omega_{\mathrm{i}}}=\frac{1-n_1 v/c}{1-n_{2+} v/c}.
\end{equation}
\end{subequations}
In these relations, the (effective) refractive indices of the second medium in the $\pm z$ directions are given by the formula~\cite{Deck_2021_electromagnetic}
\begin{equation} \label{eq:n2_pm}
    n_{2\pm}=c(\sqrt{\epsilon_{2\|}\mu_{2\|}}\pm\xi_2),
\end{equation}
with the parameters in Eqs.~\eqref{eq:para_matter} for the M system, in Eqs.~\eqref{eq:para_interface} for the I system and in Eqs.~\eqref{eq:para_K_sub} and~\eqref{eq:para_Kp_sub} for the H system.

The scattering coefficients may be directly derived in the $K$ frame from the \emph{moving boundary conditions}~\cite{Van_2012_relativity,Deck_2023_yeecell}
\begin{subequations} \label{eq:bond_cond_gene}
    \begin{equation} 
        \E_{1\|}+\mathbf{v} \times \B_{1\|}=\E_{2\|}+\mathbf{v} \times \B_{2\|}
    \end{equation}
    and
    \begin{equation} 
        \H_{1\|}-\mathbf{v} \times \D_{1\|}=\H_{2\|}-\mathbf{v} \times \D_{2\|}.
    \end{equation}
\end{subequations}
Eliminating all the field quantities, we obtain the subluminal reflection coefficient $r=A_{\mathrm{r}}/A_{\mathrm{i}}$ and transmission coefficient $\tau=A_{\mathrm{t}}/A_{\mathrm{i}}$ as
\begin{subequations} \label{eq:r_t}
    \begin{equation} \label{eq:sub_r_t}
        r = \frac{\eta_2-\eta_1}{\eta_1+\eta_2}\frac{1-n_1 v/c}{1+n_1 v/c} \quad \textrm{and} \quad
        \tau = \frac{2 \eta_2}{\eta_1+\eta_2}\frac{1-n_1 v/c}{1-n_{2+} v/c},
\end{equation}
and the superluminal reflection coefficient $\zeta=A_{\zeta}/A_{\mathrm{i}}$ and transmission coefficient $\xi=A_{\xi}/A_{\mathrm{i}}$ as
\begin{equation} \label{eq:sup_r_t}
        \zeta = \frac{\eta_1-\eta_2}{2\eta_1}\frac{1-n_1 v/c}{1+n_{2-} v/c} \quad \textrm{and} \quad
        \xi = \frac{\eta_1+\eta_2}{2\eta_1}\frac{1-n_1 v/c}{1-n_{2+} v/c},
\end{equation}
\end{subequations}
where
\begin{equation} \label{eq:etas}
    \eta_{1}=\sqrt{\frac{\mu_1}{\epsilon_1}}\quad\textrm{and}\quad\eta_{2}=\sqrt{\frac{\mu_{2\|}}{\epsilon_{2\|}}}\
\end{equation}
are the wave impedances of the two media, with the second medium parameters given in Eqs.~\eqref{eq:para_matter} for the M system, in Eqs.~\eqref{eq:para_interface} for the I system and in Eqs.~\eqref{eq:para_K_sub} and~\eqref{eq:para_Kp_sub} for the H system.

\subsection{Comparison of Scattering in the system P [Fig.~\ref{fig:illustration}(c)] and the system H [Fig.~\ref{fig:illustration}(d)]}

\begin{figure}[ht!]
    \centering
    \includegraphics[width=8.6cm]{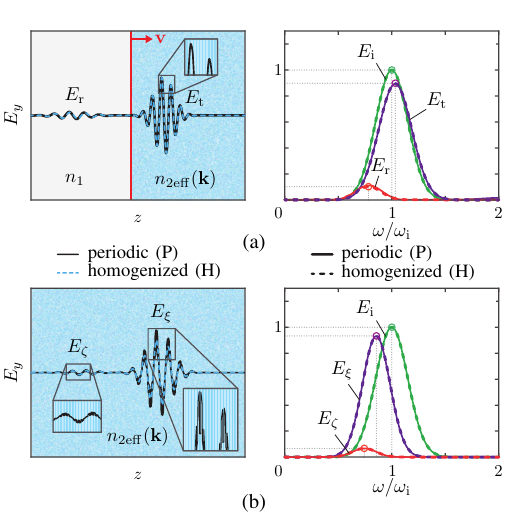}
    \caption{\label{fig:homo_periodic} Full-wave (FDTD) simulations of the scattering of a modulated Gaussian pulse at a periodic [P system, Fig.~\ref{fig:illustration}(c)] and homogenized [H system, Fig.~\ref{fig:illustration}(d)] moving-perturbation structure with $\epsilon_{\mathrm{r}1}=\epsilon_{\mathrm{r}b}=1.5$, $\epsilon_{\mathrm{r}a}=3$ and $\mu_{\mathrm{r}1}=\mu_{\mathrm{r}a}=\mu_{\mathrm{r}b}=1$, moving at the (a)~subluminal velocity, $v=0.1c$ and (b)~superluminal velocity, $v=10c$. In both cases, the unit-cell size is set to the sub-wavelength value of $l=\lambda_{\mathrm{i}}/10$ with $l_a=l_b$ [Fig.~\ref{fig:illustration}(c)]. The left panels show snapshots capturing the scattered waves after interaction with the modulation interfaces and the right panels present the Fourier transforms of the scattered and incident pulses. The insets in the left panels show enlarged waveforms within the periodic structure to better illustrate the scattering waveforms at each interface of the periodic layered structure.}
\end{figure}

Figure~\ref{fig:homo_periodic} compares the scattering of a normally incident modulated Gaussian pulse at an actual moving-perturbation periodic structure with its homogenized counterpart, based on full-wave (FDTD) simulations~\cite{Deck_2023_yeecell,Bahrami_2023_FDTD}. We consider here a subluminal case with $v=0.1c$ in Fig.~\ref{fig:homo_periodic}(a) and a superluminal case with $v=10c$ in Fig.~\ref{fig:homo_periodic}(b). The left panels illustrate the scattering phenomena at a specific time after interaction with the interfaces. Notably, the more pronounced ripples in the transmitted wave scattered by the superluminal periodic structure [see the inset in Fig.~\ref{fig:homo_periodic}(b)] arise from the larger variations in wave magnitude [Eqs.~\eqref{eq:r_t}] at the interfaces of the layers, as compared to the subluminal case [see the inset in Fig.~\ref{fig:homo_periodic}(a)]\footnote{Neglecting the minor reflections within each layer, the transmission coefficient at the interface between media $a$ and $b$ is approximately $1.7$ for the superluminal case [Eq.~\eqref{eq:sup_r_t}] shown in Fig.~\ref{fig:homo_periodic}(b), while it is only around $1.1$ for the subluminal case [Eq.~\eqref{eq:sub_r_t}] shown in Fig.~\ref{fig:homo_periodic}(a). This suggests that more pronounced wave variations and ripples occur at each interface in the superluminal periodic structure compared to the subluminal structure.}. 
The right panels provide quantitative validations in terms of the Fourier-transformed field distributions, where the circles correspond to the analytical results [Eqs.~\eqref{eq:w} and~\eqref{eq:r_t}]\footnote{Note that the scattering coefficients for modulated Gaussian pulse incidence in the frequency domain are equal to those in the time domain, $r,\tau,\zeta,\xi$ [Eqs.~\eqref{eq:r_t}], divided by the pulse compansion factors, $\alpha_{\r,\t,\zeta,\xi}=\omega_{\r,\t,\zeta,\xi}/\omega_{\mathrm{i}}$, provided in Eqs.~\eqref{eq:w}, respectively~\cite{Caloz_2019_spacetime2,Deck_2023_yeecell}.}. As shown in Fig.~\ref{fig:homo_periodic}, despite the homogenization approximation considered in the derivations of the analytical results for the moving-perturbation periodic structure, the results exhibit a good agreement with the actual scattering and therefore we approximate the P system using the H system in this paper.
%

\section{Derivations of Eqs.~\eqref{eq:p_cw} and~\eqref{eq:f_cw}} \label{app:p_f}
\subsection{Derivations of Eqs.~\eqref{eq:p_cw}}
\subsubsection{Subluminal regime}
In the subluminal regime, depicted in Fig.~\ref{fig:moving_cylinder}(a), the incident and reflected waves are located on the $-z$ side of the discontinuity, while the transmitted wave is on the $+z$ side. Substituting the time-averaged Poynting vector magnitude $\langle S \rangle=A^2/(2\eta)$ and wave energy density $\langle W \rangle=A^2 n/(2\eta c)$, where $A$ is the wave amplitude and $c$ is the speed of light in vacuum, into Eq.~\eqref{eq:power_vm} yields
        \begin{equation}\label{eq:power_vm_sub}
    \begin{aligned}
        &\langle p_{\mathrm{s}}^{\mathrm{sub}\pm} \rangle= \hat{\mathbf{z}} \cdot(\langle \S_{\t} \rangle-\langle \S_{\r} \rangle-\langle \S_{\mathrm{i}} \rangle)-v (\langle W_{\t} \rangle-\langle W_{\r} \rangle-\langle W_{\mathrm{i}} \rangle) \\
        &=\frac{1}{2}\left(\frac{A^2_{\t}}{\eta_2}+\frac{A^2_{\r}}{\eta_1}-\frac{A^2_{\mathrm{i}}}{\eta_1}\right)-\frac{v}{2c}\left(\frac{n_{2+}}{\eta_2}A^2_{\t}-\frac{n_{1}}{\eta_1}A^2_{\r}-\frac{n_{1}}{\eta_1}A^2_{\mathrm{i}}\right) \\
        &=I_{\mathrm{i}}\left[\frac{\eta_1}{\eta_2}\left(1-\frac{n_{2+} v}{c}\right)\tau^2+\left(1+\frac{n_{1} v}{c}\right)r^2 -\left(1-\frac{n_{1} v}{c}\right)\right].
    \end{aligned}
    \end{equation}
\subsubsection{Superluminal regime}
In the comoving (incident wave propagating in the same direction as the interface) superluminal regime, depicted in Fig.~\ref{fig:moving_cylinder}(b), the incident wave is located on the $+z$ side of the discontinuity, while the scattered waves are on the $-z$ side. Then, Eq.~\eqref{eq:power_vm} becomes
    \begin{equation}\label{eq:power_vm_sup_co}
        \begin{split}
        &\langle p_{\mathrm{s}}^{\mathrm{sup}+} \rangle=
        \begin{aligned}[t]
        \hat{\mathbf{z}} \cdot(\langle\mathbf{S}_{\mathrm{i}}\rangle-\langle\mathbf{S}_{\xi}\rangle-\langle&\mathbf{S}_{\zeta}\rangle)\\
        &-v(\langle W_{\mathrm{i}}\rangle-\langle W_{\xi}\rangle-\langle W_{\zeta}\rangle)
        \end{aligned}
        \\
        &=\frac{1}{2}\left(\frac{A^2_{\mathrm{i}}}{\eta_1} -\frac{A^2_{\xi}}{\eta_2}+\frac{A^2_{\zeta}}{\eta_2}\right)-\frac{v}{2c}\left(\frac{n_{1}}{\eta_1}A^2_{\mathrm{i}}-\frac{ n_{2+}}{\eta_2}A^2_{\xi}-\frac{n_{2-}}{\eta_2}A^2_{\zeta}\right) \\
        &=
        \begin{aligned}[t]
            -I_{\mathrm{i}}\Bigg\{\frac{\eta_1}{\eta_2}\Bigg[\Bigg(1-\frac{n_{2+} v}{c}\Bigg)\xi^2 - \Bigg(1+&\frac{n_{2-} v}{c}\Bigg)\zeta^2\Bigg] \\
        &-\Bigg(1-\frac{n_{1} v}{c}\Bigg)\Bigg\}.
        \end{aligned}
        \end{split}
    \end{equation}
    
In the contramoving (incident wave propagating in the opposite direction of the interface) superluminal regime, the incident wave is located on the $-z$ side of the discontinuity, while the scattered waves are on the $+z$ side. We have then
    \begin{equation}\label{eq:power_vm_sup_contra}
        \begin{split}
        &\langle p_{\mathrm{s}}^{\mathrm{sup}-} \rangle=\hat{\mathbf{z}} \cdot(\langle \mathbf{S}_{\xi}\rangle+ \langle \mathbf{S}_{\zeta}\rangle-\langle \mathbf{S}_{\mathrm{i}}\rangle)-v(\langle W_{\xi}\rangle+\langle W_{\zeta}\rangle-\langle W_{\mathrm{i}}\rangle) \\
        &=\frac{1}{2}\left(\frac{A^2_{\xi}}{\eta_2}-\frac{A^2_{\zeta}}{\eta_2}-\frac{A^2_{\mathrm{i}}}{\eta_1}\right) -\frac{v}{2c}\left(\frac{n_{2+}}{\eta_2}A^2_{\xi}+\frac{n_{2-}}{\eta_2}A^2_{\zeta}-\frac{n_{1}}{\eta_1}A^2_{\mathrm{i}}\right)\\
        &=
        \begin{aligned}[t]
            I_{\mathrm{i}}\Bigg\{\frac{\eta_1}{\eta_2}\Bigg[\Bigg(1-\frac{n_{2+} v}{c}\Bigg)\xi^2 - \Bigg(1+&\frac{n_{2-} v}{c}\Bigg)\zeta^2\Bigg] \\&-\Bigg(1-\frac{n_{1} v}{c}\Bigg)\Bigg\}.
        \end{aligned}
    \end{split}
    \end{equation}
\subsection{Derivations of Eqs.~\eqref{eq:f_cw}} \label{app:f}
\subsubsection{Subluminal regime}
Substituting the time-averaged Maxwell stress tensor component $\langle T_{zz}\rangle=A^2 n/(2\eta c)$ and wave momentum density magnitude $\langle g \rangle=A^2n^2/(2\eta c^2)$ into Eq.~\eqref{eq:f_vm}, we obtain
    \begin{equation}\label{eq:f_vm_sub}
        \begin{split}
        &\langle f_{\mathrm{s}}^{\mathrm{sub}\pm}\rangle=
        (\langle T_{zz\t}\rangle-\langle T_{zz\r}\rangle-\langle T_{zz\mathrm{i}}\rangle)
        -v(\langle g_{\t}\rangle+\langle g_{\r}\rangle-\langle g_{\mathrm{i}}\rangle)\\
        &=
        \begin{aligned}[t]
            \frac{1}{2c}\Bigg(\frac{n_{2+}}{\eta_2}A^2_{\t}-\frac{n_{1}}{\eta_1}A^2_{\r}&-\frac{n_{1}}{\eta_1}A^2_{\mathrm{i}}\Bigg) \\ & - \frac{v}{2c^2}\Bigg(\frac{n_{2+}^{2}}{\eta_{2}}A_{\t}^2+\frac{n_{1}^{2}}{\eta_{1}}A_{\r}^2-\frac{n_{1}^{2}}{\eta_{1}}A_{\mathrm{i}}^2 \Bigg)
        \end{aligned}
         \\
        &=
        \begin{aligned}[t]
        \frac{I_{\mathrm{i}}}{c}\Bigg[\frac{\eta_1 n_{2+}}{\eta_2 }\Bigg(1-\frac{n_{2+} v}{c}\Bigg)\tau^2-n_1\Bigg(1+&\frac{n_{1} v}{c}\Bigg)r^2\\ &-n_1\Bigg(1-\frac{n_{1} v}{c}\Bigg)\Bigg],
        \end{aligned}
        \\
        &=
        \begin{aligned}[t]
        \frac{I_{\mathrm{i}}}{v}\Bigg[\frac{\eta_1}{\eta_2}\Bigg(1-\frac{n_{2+} v}{c}\Bigg)\tau^2 +\Bigg(1+\frac{n_{1} v}{c}\Bigg)r^2 -\Bigg(1-\frac{n_{1} v}{c}\Bigg)\Bigg].
        \end{aligned}
    \end{split}
    \end{equation}
    where the last equality is obtained by substituting Eqs.~\eqref{eq:sub_r_t} and then rewriting the result in terms of $\tau$ and $r$.
\subsubsection{Superluminal regime}
Similarly, in the superluminal regime, Eq.~\eqref{eq:f_vm} becomes
\begin{equation}\label{eq:f_vm_sup_co}
    \begin{split}
        &\langle f_{\mathrm{s}}^{\mathrm{sup}\pm}\rangle=
        \begin{aligned}[t]
        \pm(\langle T_{zz\mathrm{i}}\rangle-\langle T_{zz\xi}\rangle-&\langle T_{zz\zeta}\rangle)
        \\
        &\mp v(\langle g_{\mathrm{i}}\rangle-\langle g_{\xi}\rangle+\langle g_{\zeta}\rangle)
        \end{aligned}
        \\
        &=
        \begin{aligned}[t]
            \pm \frac{1}{2c}\Bigg(\frac{ n_1}{\eta_1}A^2_{\mathrm{i}} -&\frac{n_{2+}}{\eta_2}A^2_{\xi}-\frac{n_{2-}}{\eta_2}A^2_{\zeta}\Bigg) \\ &
            \mp \frac{v}{2c^2}\Bigg(\frac{n_{1}^{2}}{\eta_1}A^2_{\mathrm{i}}-\frac{n_{2+}^{2}}{\eta_2}A^2_{\xi}+\frac{n_{2-}^{2}}{\eta_2}A^2_{\zeta}\Bigg)
        \end{aligned}
         \\
        &=
        \begin{aligned}[t]
            \mp\frac{I_{\mathrm{i}}}{v}\Bigg\{\frac{\eta_1}{\eta_2}\Bigg[\Bigg(1-\frac{n_{2+} v}{c}\Bigg)\xi^2 - \Bigg(1+&\frac{n_{2-} v}{c}\Bigg)\zeta^2\Bigg] \\
        &-\Bigg(1-\frac{n_{1} v}{c}\Bigg)\Bigg\}.
        \end{aligned}
    \end{split}
\end{equation}

{\small
\bibliographystyle{IEEEtran}
\bibliography{Reference.bib}
}

\end{document}